\documentclass[a4paper,11pt]{article}
\usepackage{authblk}
\usepackage{hyperref}
\usepackage{color}

\title{\textbf{A Theory of Theories}}

\author{Mich\`ele Levi}
\affil{Mathematical Institute, University of Oxford, \\ 
	Oxford OX2 6GG, United Kingdom\\
	\bigskip
	levi@maths.ox.ac.uk}
\date{}

\begin{document}

\maketitle

\begin{abstract}
We take a tour through the past, present and future 
of Effective Field Theory, 
with applications ranging 
from LHC physics to cosmology.	
\end{abstract}

\clearpage


High-energy physics spans a wide range of energies,
from a few MeV to TeV, that are all relevant. It is
therefore often difficult to take all phenomena into
account at the same time. Effective field theories (EFTs) are
designed to break down this range of scales into smaller
segments so that physicists can work in the relevant range.
Theorists ``cut'' their theory's energy scale at the order of
the mass of the lightest particle omitted from the theory,
such as the proton mass. Thus, multi-scale problems reduce
to separate and single-scale problems.
EFTs are today also understood to be ``bottom-up''
theories. Built only out of the general field content and
symmetries at the relevant scales, they allow us to test
hypotheses efficiently and to select the most promising
ones without needing to know the underlying theories
in full detail. Thanks to their applicability to all generic
classical and quantum field theories, the sheer variety of
EFT applications is striking.

In hindsight, particle physicists were working with
EFTs from as early as Fermi's phenomenological picture
of beta decay in which a four-fermion vertex replaces the
W-boson propagator because the momentum is much
smaller compared to the mass of the W boson.
Like so many profound concepts in
theoretical physics, EFT was first considered in a narrow
phenomenological context. One of the earliest instances
was in the 1960s, when ad-hoc methods of current algebras
were utilised to study weak interactions of hadrons. This
required detailed calculations, and a simpler approach
was needed to derive useful results. The heuristic idea of
describing hadron dynamics with the most general Lagrangian 
density based on symmetries, the relevant energy
scale and the relevant particles, which can be written in
terms of operators multiplied by Wilson coefficients, was
yet to be known. With this approach, it was possible to
encode local symmetries in terms of the current algebra
due to their association with conserved currents.

For strong interactions, physicists described the 
interaction between pions with chiral perturbation theory, 
an effective Lagrangian, which simplified current algebra
calculations and enabled the low-energy theory to be
investigated systematically. This ``mother'' of modern
EFTs describes the physics of hadrons and remains valid
to an energy scale of the proton mass. Heavy-quark 
effective theory (HQET), introduced by Howard Georgi in 1990,
complements chiral perturbation theory by describing the
interactions of charm and bottom quarks. HQET allowed
us to make predictions on B-meson decay rates, since
the corrections could now be classified. The more powers
of energy are allowed, the more infinities appear. These
infinities are cancelled by available counter-terms.

Similarly, it is possible to regard the Standard Model as
the truncation of a much more general theory including
non-renormalizable interactions, which yield corrections
of higher order in energy. This perception of the whole
Standard Model as an effective field theory started to be
formed in the late 1970s by Weinberg and others. 
Among
the known corrections to the Standard Model that do not
satisfy its approximate symmetries are neutrino masses,
postulated in the 1960s and discovered via the observation
of neutrino oscillations in the late 1990s. While the scope
of EFTs was unclear initially, today we understand that
all successful field theories, with which we have been
working in many areas of theoretical physics, are nothing
but effective field theories. EFTs provide the theoretical
framework to probe new physics and to establish precision
programmes at experiments. The former is crucial for
making accurate theoretical predictions, while the latter
is central to the physics programme of CERN in general.

\section{EFTs in Particle Physics}

More than a decade has passed since the first run of the
LHC, in which the Higgs boson and the mechanism for
electroweak symmetry breaking were discovered. So far,
there are no signals of new physics beyond the SM. EFTs
are well suited to explore LHC physics in depth. A typical
example for an event involving two scales is Higgs-boson
production because there is a factor $10-100$ between its
mass and transverse momentum. The calculation of
each Higgs-boson production process leads to large logarithms 
that can invalidate perturbation theory due to the
large-scale separation. This is just one of many examples
of the two-scale problem that arises when the full quantum
field theory approach for high-energy colliders is applied.
Traditionally, such two-scale problems have been treated
in the framework of QCD factorisation and resummation.

Over the past two decades, it has been possible to recast
two-scale problems at high-energy colliders with the
advent of soft-collinear effective theory (SCET). SCET is
nowadays a popular framework that is used to describe
Higgs physics, jets and their substructure, as well as more
formal problems, such as power corrections to reconstruct 
full amplitudes eventually. The difference between
HQET and SCET is that SCET considers long-distance
interactions between quarks and both soft and collinear
particles, whereas HQET takes into account only soft
interactions between a heavy quark and a parton. SCET
is just one example where the EFT methodology has been
indispensable, even though the underlying theory at much
higher energies is known. Other examples of EFT 
applications include precision measurements of rare decays
that can be described by QCD with its approximate chiral
symmetry, or heavy quarks at finite temperature and
density. EFT is also central to a deeper understanding of
the so-called flavour anomalies, enabling comparisons
between theory and experiment in terms of particular
Wilson coefficients.

Moreover, precision measurements of Higgs and electroweak 
observables at the LHC and future colliders will
provide opportunities to detect new physics signals, such
as resonances in invariant mass plots, or small deviations
from the SM, seen in tails of distributions for instance at the
HL-LHC -- testing the perception of the SM as a low-energy
incarnation of a more fundamental theory ­­being probed at
the electroweak scale. This is dubbed the SMEFT (SM EFT)
or HEFT (Higgs EFT), depending on whether the Higgs fields
are expressed in terms of the Higgs doublet or the physical
Higgs boson. This particular EFT framework has recently
been implemented in the data-analysis tools at the LHC,
enabling the analyses across different channels and even
different experiments.
At the same time, the study of SMEFT and HEFT has sparked a
plethora of theoretical investigations that have uncovered
its remarkable underlying features, for example allowing
EFT to be extended or placing constraints on the EFT coefficients 
due to Lorentz invariance, causality and analyticity.

\section{EFTs in Gravity}

Since the inception of EFT, it was believed that the framework 
is applicable only to the description of quantum field
theories for capturing the physics of elementary particles
at high-energy scales, or alternatively at very small length
scales. Thus, EFT seemed mostly irrelevant regarding gravitation, 
for which we are still lacking a full theory valid at
quantum scales. The only way in which EFT seemed to be
pertinent for gravitation was to think of general relativity
as a first approximation to an EFT description of quantum
gravity, which indeed provided a new EFT perspective at
the time. However, in the past decade it has become widely
acknowledged that EFT provides a powerful framework to
capture gravitation occurring completely across large length
scales, as long as these scales display a clear hierarchy.

The most notable application to such classical gravitational 
systems came when it was realised that the EFT
framework would be ideal to handle gravitational radiation 
emitted at the inspiral phase of a binary of compact
objects, such as black holes. At this phase in the evolution 
of the binary, the compact objects are moving at
non-relativistic velocities. Using the small velocity as the
expansion parameter exhibits the separation between the
various characteristic length scales of the system. Thus, the
physics can be treated perturbatively. For example, it was
found that even couplings manifestly change in classical
systems across their characteristic scales, which was 
previously believed to be unique to quantum field theories. 
The application of EFT to the binary inspiral problem has been
so successful that the precision frontier has been pushed
beyond the state of the art, quickly surpassing the reach
of work that has been focused on the two-body problem
for decades via traditional methods in general relativity.

This theoretical progress has made an even broader impact 
since the breakthrough direct discovery of gravitational 
waves (GWs) was announced in 2016. An inspiraling
binary of black holes merged into a single black hole in
less than a split second, releasing an enormous amount of
energy in the form of GWs, which instigated even greater,
more intense use of EFTs for the generation of theoretical
GW data. In the coming years and decades, a continuous
increase in the quantity and quality of real-world GW data
is expected from the rapidly growing worldwide network
of ground-based GW detectors, and future space-based
interferometers, covering a wide range of target frequencies.

\section{EFTs in Cosmology}

Cosmology is inherently a cross-cutting domain, spanning
scales over about $10^{60}$ orders of magnitude, from the Planck
scale to the size of the observable universe. As such, 
cosmology generally cannot be expected to be tackled directly
by each of the fundamental theories that capture particle
physics or gravity. The correct description of cosmology
relies heavily on the work in many disparate areas of research
in theoretical and experimental physics, including particle
physics and general relativity among many more.

The development of EFT applications in cosmology --
including EFTs of inflation, dark matter, dark energy and
even EFTs of large-scale structure -- has become essential
to make observable predictions in cosmology. The discovery 
of the accelerated expansion of the universe in 1998
shows our difficulty in understanding gravity both at the
quantum regime and the classical one. The cosmological
constant problem and dark-matter paradigm might be a
hint for alternative theories of gravity at very large scales.
Indeed, the problems with gravity in the very-high and
very-low energy range may well be tied together. The science
programme of next-generation large surveys, such as ESA's
Euclid satellite, 
rely heavily on all these EFT applications for the exploitation 
of the enormous data that is going to be collected to constrain
unknown cosmological parameters, thus helping to pinpoint
viable theories.

\section{The Future of EFTs in Physics}

The EFT framework plays a key role at the exciting and
rich interface between theory and experiment in particle
physics, gravity and cosmology as well as in other domains,
such as condensed-matter physics, which were not covered
here. The technology for precision measurements in these
domains is constantly being upgraded, and in the coming
years and decades we are heading towards a growing influx
of real-world data of higher quality. Future particle-collider
projects, such as the Future Circular Collider at CERN, or
China's Circular Electron Positron Collider, are being planned
and developed. Precision cosmology is also thriving, with
an upcoming next-generation of very large surveys, such as
the ground-based LSST, or space-based Euclid. GW detectors
keep improving and multiplying, and besides those that are
currently operating many more are planned, aimed at measuring 
various frequency ranges, which will enable a richer
array of sources and events to be found.

Half a century after the concept has formally emerged,
effective field theory is still full of surprises. Recently, the
physical space of EFTs has been studied as a fundamental
entity in its own right. These studies, by numerous groups
worldwide, have exposed a new hidden ``totally positive''
geometric structure dubbed the EFT-hedron that constrains
the EFT expansion in any quantum field theory, and even
string theory, from first principles, including causality,
unitarity and analyticity, to be satisfied by any amplitudes
of these theories. This recent formal progress reflects the
ultimate leap in the perception of EFT nowadays as the most
fundamental and most generic theory concept to capture
the physics of nature at all scales. Clearly, in the vast array
of formidable open questions in physics that still lie ahead,
effective field theory is here to stay -- for good.

\section*{Acknowledgements}

We dedicate this article to the memory of Steven Weinberg, who so generously graced us
with a spectacular inaugural lecture to the international online series hosted at CERN 
``All Things EFT'', which turned out to be his final published lecture.

\noindent We thank Cliff Burgess and HuaXing Zhu for comments and input on a preliminary
draft. 
\noindent ML has been supported by the Science and Technology Facilities Council (STFC) 
Rutherford Grant ST/V003895 
``\textit{Harnessing QFT for Gravity}'', 
and by the Mathematical Institute University of Oxford.


\end{document}